\documentclass[10pt,letterpaper]{article}
\usepackage{opex3}
\usepackage{color}

\usepackage{caption}
\usepackage{verbatim}
\usepackage{subfig}
\usepackage{amsmath}
\usepackage[toc,page]{appendix}
\usepackage{epstopdf}
\usepackage{cite}
\usepackage{caption}
\usepackage{subfig}

\begin{document}
\title{A temporal model for quasi-phase matching in high-order harmonic generation}

\author{Y. Tao$^{1,*}$, S. J. Goh$^1$, H. M. J. Bastiaens$^1$, P. J. M. van der Slot$^1$, S. G. Biedron$^2$, S. V. Milton$^2$ and  K. -J. Boller$^1$}

\address{$^1$Laser Physics and Nonlinear Optics, Mesa+ Institute for Nanotechnology, University of Twente, Enschede, The Netherlands\\
$^2$ Department of Electrical and Computer Engineering, Colorado State University, Fort Collins, Colorado, USA}

\email{$^*$y.tao-1@utwente.nl} 



\begin{abstract}
We present a model for quasi-phase matching (QPM) in high-order harmonic generation (HHG). Using a one-dimensional description, we analyze the time-dependent, ultrafast wave-vector balance to calculate the on-axis harmonic output versus time, from which we obtain the output pulse energy. Considering, as an example, periodically patterned argon gas, as may be provided with a grid in a cluster jet, we calculate the harmonic output during different time intervals within the drive laser pulse duration. We find that identifying a suitable single spatial period is not straightforward due to the complex and ultrafast plasma dynamics that underlies HHG at increased intensities. The maximum on-axis harmonic pulse energy is obtained when choosing the QPM period to phase match HHG at the leading edge of the drive laser pulse.
\end{abstract}



\section{Introduction}
\label{sec: intro}
%
%
High-order harmonic generation (HHG) is an extremely nonlinear optical process that provides coherent radiation in the extreme ultraviolet spectral region on an ultrafast time scale. Such radiation is of great interest for applications such as coherent diffractive imaging of nanostructures~\cite{Witte2014}, pump-probe spectroscopy for studying dynamical electronic processes inside molecules~\cite{Mairesse2010} and attosecond science~\cite{Baker2006}. The range of generated wavelengths covered by the harmonics is usually limited, on the long wavelength side by reabsorption in the generating medium, and on the short wavelength side by lack of phase matching due to ionization. For example, the maximum harmonic order (the shortest wavelength) which can be phase matched in argon is the 35$^{\rm{th}}$ order ($\approx$24 nm) when using femtosecond pulses from an 800 nm drive laser~\cite{Popmintchev2009}. The laser intensities required for generating radiation of even shorter wavelengths are so high that a significant fraction of the gaseous medium becomes ionized. As a result of the generated transient plasma, there is an additional refractive index contribution which renders phase matching techniques ineffective due to wave-vector mismatch.

A technique that is generally seen as suitable for overcoming this ionization limitation and that allows one to efficiently generate higher order harmonics is so-called \emph{quasi-phase matching} (QPM)~\cite{Popmintchev2009}. With the introduction of a spatially periodic structure of suitable periodicity, an extra wave vector can be provided that compensates the wave-vector mismatch. Various schemes to demonstrate or make use of QPM in HHG have been proposed and developed. The idea is that efficient HHG is only enabled in the in-phase region and is suppressed in the out-of-phase region. As an example, a train of tunable counter-propagating pulses in a hollow-core waveguide has been used to introduce a spatially periodic ionization pattern for the generation of photon energies of several hundred electron volts~\cite{Zhang2007}. Other schemes include waveguides with a spatially periodic diameter~\cite{Gibson2003}, multimode beating in a capillary~\cite{Zepf2007} and polarization-beating in a birefringent waveguide~\cite{Liu2012, Liu2013}. Recently, a small set of periodically spaced dual-gas jets, consisting of the generating gas (argon) and a passive gas (hydrogen), has shown to shift harmonic generation up to the 41$^{\rm{st}}$ order ($\approx$19 nm)~\cite{Hage2014, Willner2011PRL, Willner2011NJP}. 

All these approaches are based on the assumption that, depending on the specific set of parameters used, there is a certain spatial period, $P_{mod}$, that provides phase matching. This assumption is much in analogy with second-order nonlinear conversion in periodically poled crystals where a particular conversion process with given input and output wavelengths requires a well-defined poling period. However, selecting an appropriate periodic structure for achieving QPM in HHG is actually rather problematic, since the wave-vector mismatch is strongly time dependent throughout the entire drive laser pulse. A fixed periodic structure may only be effective in achieving QPM during a certain short time interval within the drive laser pulse, such that the time-integrated (total) output pulse energy may still remain low. To demonstrate QPM for HHG in experiments using the schemes mentioned above, the selection of a fixed-periodic structure for maximizing the total high-order harmonic (HH) output is usually based on an unproven assumption, namely, that the quasi-phase matching period has to be chosen equal to twice the coherence length calculated for the time-instance where the intensity of the drive laser pulse is maximum (at peak intensity). This is based on the argument that the nonlinearly induced dipole moment of the individual atoms is the highest at that time~\cite{Lewenstein1994}. However, this assumption neglects any radiation that may be generated before and after the peak of the drive laser pulse. A direct experimental clarification is not known to us and difficult to carry out, due to the ultrashort time scales. Instead, one is usually detecting or optimizing any phase matching in HHG via the output pulse energy. Thus, for a proper discussion of optimizing QPM in HHG, one requires a time-dependent model of the output followed by a calculation of the output pulse energy.

Here, we present the first time-dependent model for quasi-phase matching in high-order harmonic generation. We call this an extended model because it combines previously known physical effects into a more encompassing theory that delivers somewhat deviating predictions for optimizing the harmonic output. Using a one-dimensional model, we include the full wave-vector mismatch with its ultrafast time dependence to calculate the HH field vs. time. The pulse energy is then obtained by time integration. For definiteness, and as a typical example, we consider HHG in argon gas where a periodic density pattern is provided, e.g., with a grid in a supersonic jet. We show that the maximum harmonic pulse energy is obtained when choosing the QPM period for phase matching in the leading edge of the drive laser pulse, rather than at the peak.
\section{One-dimensional time-dependent model for HHG}
\label{sec: one-dimensional model}
As our basis for investigating quasi-phase matching, we recall the well-known model of Constant et al.~\cite{Constant1999} and Kazamias et al.~\cite{Kazamias2003-2}, which describes standard phase matching as achievable via pressure adjustment at weakly ionizing intensities.
The instantaneous field for a specific harmonic order, $q$, generated at the exit of the medium, $E_{q}(t)$, can be expressed as
\begin{equation}
\label{eq: instantaneous HHG field}
E_{q}(t)\propto\int^{L_{med}}_{0}\rho(z)\mathrm{exp}\left(-\frac{L_{med}-z}{2L_{abs}(z)}\right)d_q(z,t)\left[1-\eta(t)\right]\mathrm{exp}\left[i\phi_q(z,t)\right]dz.
\end{equation}
Here, $L_{med}$ is the length of the medium, $z$ is the propagation coordinate, $\eta(t)$ is the ionization fraction, $\rho(z)$ is the initial atomic number density, $L_{abs}(z)=1/\left[\sigma\rho(z)\right]$ is the absorption length, where $\sigma$ is the ionization cross section~\cite{Chan1992}, and $\phi_q(z,t)$ is the accumulated phase difference between the drive laser field and the generated harmonic field at the exit of the medium. The drive laser field is implicitly included in the induced atomic dipole moment amplitude, $d_q(z,t)$. Based on the full quantum theory developed by Lewenstein et al.~\cite{Lewenstein1994}, $d_q$ exhibits a strong dependence on time through its variation with the drive laser intensity, $I(z,t)$, and can be described approximately by the following power law~\cite{kazamias2011}.
\begin{equation}
\label{eq: Aq}
d_q(z,t)\left\{
\begin{aligned}
\propto\left(\frac{I(z,t)}{I_{thr}(q)}\right)^{4.6} \qquad &\textrm{if} \qquad I>I_{thr}(q) \\
\approx0\qquad \qquad &\textrm{if} \qquad I<I_{thr}(q)
\end{aligned}
\right.
\end{equation}
Here, $I_{thr}(q)$ is the threshold laser intensity to be reached before there is a noticeable generation of the $q^{\rm{th}}$ harmonic. The factor $1-\eta(t)$ in Eq. \eqref{eq: instantaneous HHG field} originates from ionization-induced ground state depletion, which indicates that the generated harmonic field scales with the number of neutral atoms. The ionization fraction, $\eta(t)$, can be calculated using the Ammosov-Delone-Krainov (ADK) model~\cite{Ammosov1986}. 
\begin{figure}[ht!]
\centering
\includegraphics[width=0.8\textwidth]{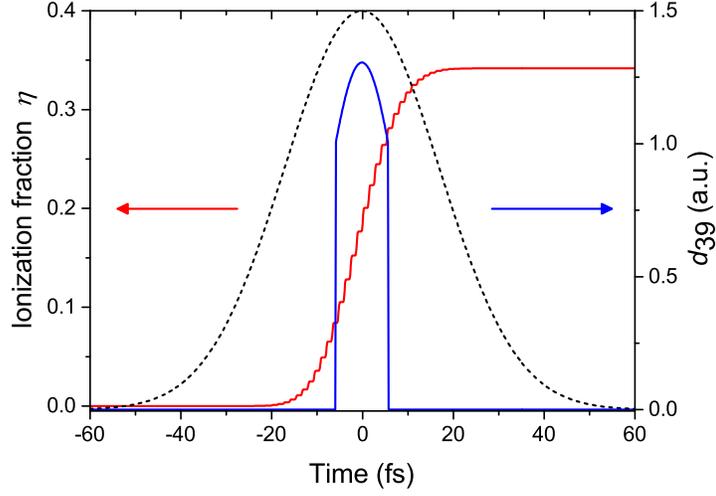}
\caption{The ionization fraction, $\eta$, (red line) and the harmonic dipole amplitude, $d_{39}$, for the 39$^{\rm{th}}$ harmonic ($\approx20$ nm, blue line) calculated for argon and a drive laser pulse (Gaussian temporal shape) with a central wavelength at 800 nm, a pulse duration of $\tau=40$ fs (FWHM) and a peak intensity of $I_0=2.5\times{10^{14}\;\rm{W/cm^2}}$. The drive laser pulse envelop (black dashed line) is plotted for reference.}
\label{fig: ionization_fraction}
\end{figure}

To give an example for the typical temporal variation of the ionization fraction and atomic harmonic dipole amplitude, Fig. \ref{fig: ionization_fraction} shows
a calculation of the ionization fraction and the 39$^{\rm{th}}$ harmonic ($\approx20$ nm) dipole amplitude, $d_{\mathrm{39}}$, of argon for a drive laser pulse with a peak intensity of $I_0=2.5\times{10^{14}\;\rm{W/cm^2}}$. The laser pulse has a Gaussian shape with a pulse duration of $\tau=40$ fs (FWHM) and a central laser wavelength of 800 nm. As can be seen from the red curve, the ionization fraction reaches a value of about 18\% at the peak of the laser intensity and raises further to about 34\% towards the end of the laser pulse. It can further be noticed that the ionization fraction grows with each optical half-cycle resembling a stepwise increase (here 30 steps in total). The dipole amplitude exhibits a strong dependence on the drive laser intensity through the threshold and power law in Eq. \eqref{eq: Aq}. 

The term $\phi_q(z,t)$ in Eq. \eqref{eq: instantaneous HHG field}, is directly related to the wave-vector mismatch, $\Delta{k}(z,t)$, between the drive laser and the generated harmonics. The accumulated phase difference after a propagation length of $L_{med}$ can be described as
\begin{equation}
\phi_q(L_{med},t)=\int_{0}^{L_{med}}\Delta{k}(z,t)dz.
\label{eq: phase difference}
\end{equation}
Here, as will be detailed below, the mismatch, $\Delta{k}(z,t)$, originates from the dispersion due to neutral gas atoms ($\Delta{k_{atom}}$), electrons in the plasma that are generated via ionization ($\Delta{k_{plasma}}$), the Gouy phase shift, which depends on the focusing geometry of the drive laser ($\Delta{k}_{geo}$), and the intrinsic harmonic dipole phase ($\Delta{k_{dipole}}$), and can be written as follows~\cite{Dachraoui2008}:
\begin{equation}
\label{eq: total phase mismatch}
\Delta{k}(z,t)=\Delta{k_{atom}}(z,t)+\Delta{k_{plasma}}(z,t)+\Delta{k_{geo}}(z)+\Delta{k_{dipole}}(z,t).
\end{equation}

The contribution of the neutral atomic dispersion to the total wave-vector mismatch is positive and can be expressed by
\begin{equation}
\label{eq: neutral atomic dispersion time dependent}
\Delta{k_{atom}}(z,t)=q\left[1-\eta(z,t)\right]\frac{2\pi{P(z)}}{\lambda}\Delta{n}.
\end{equation}
Here, $P(z)$ is the pressure of the gas in mbar, $\lambda$ is the wavelength of the drive laser and $\Delta{n}$ is the difference between the indices of refraction of the gas per mbar at the drive laser wavelength, $\lambda$, and the high harmonic wavelength, $\lambda/q$. The dependence on $\eta(z,t)$ shows that the atomic dispersion decreases during the drive laser pulse.

The contribution due to the free electrons in the plasma, on the contrary, is negative and becomes
\begin{equation}
\label{eq: plasma dispersion time dependent}
\Delta{k_{plasma}}(z,t)=-\eta(z,t)P(z)N_{a}r_{e}\frac{q^2-1}{q}\lambda,
\end{equation}
where $N_a$ is the initial number density of atoms per mbar, and $r_e$ is the classical radius of the electron. This contribution via $\eta(z,t)$ grows in magnitude vs. time.

The wave-vector mismatch that arises from the Gouy phase shift, resulting from the focusing of the drive laser beam, only varies in space. It also has a negative sign and can be expressed by
\begin{equation}
\label{eq: Geometrical dispersion time dependent}
\Delta{k_{geo}}(z)=-\frac{1}{z_R+\frac{(z-z_0)^2}{z_R}}(q-1),
\end{equation}
where $z_0$ is the position of the focus and $z_R$ is the Rayleigh range. 

Finally, the contribution of the intrinsic harmonic dipole phase originates from the intensity dependence of the trajectories that the transiently ionized electrons follow before recombination and harmonic generation, which can be described as~\cite{Auguste2007}
\begin{equation}
\label{eq: atomic dipole phase dispersion time dependent}
\Delta{k_{dipole}}(z,t)=-\alpha_q\frac{\partial{I(z,t)}}{\partial{z}}=\frac{2(z-z_0)}{z_R^2\left[1+\left(\frac{(z-z_0)}{z_R}\right)^2\right]^2}\alpha_qI_0,
\end{equation}
where $\alpha_q$ is a coefficient related to the trajectories which the electrons take.
This expression can be chosen to be either positive or negative depending on the position of the focus with regard to the generating medium. Placing the medium in front of (behind) the focus, where $z<z_0$ ($z>z_0$), results in a negative (positive) term. When the focus is placed in the center of the medium ($z=z_0$), such as to maximize the drive laser intensity, the value of $\Delta{k_{dipole}}(z,t)$ becomes equal to zero.
 
The terms described so far can now be used to calculate the spatial and temporal evolution of the harmonic field, as expressed in Eq. \eqref{eq: instantaneous HHG field}.
The instantaneous harmonic intensity of the $q^{\rm{th}}$ harmonic at the exit of the medium is obtained from $I_{q}(t)=\frac{1}{2}c\varepsilon_{0}|E_{q}(t)|^2$, where $c$ and $\varepsilon_{0}$ are the speed of light and the vacuum permittivity respectively. The on-axis harmonic pulse energy of the $q^{\rm{th}}$ harmonic, $F_q$, can be obtained by temporal integration over the entire pulse and multiplying with a cross-sectional area, $S$,
\begin{equation}
\label{eq: HHG intensity}
F_q={S}\int_{-\infty}^{\infty}I_{q}(t)dt.
\end{equation}

The described model has been widely and successfully applied to analyze measurements of HHG under various experimental conditions such as guided-wave geometries~\cite{Durfee1999, Popmintchev2009}, or loose focusing geometries~\cite{Constant1999, Hergott2002, Kazamias2003, Midorikawa2008}. In the next section, we will use the model in its simplest form, namely, for the case of a homogeneous medium (argon) with a loosely focused drive laser beam. In detail, we will present the consequence of the time dependence in the wave-vector mismatch.

\section{Time-dependence of phase matching in a homogeneous medium}
\label{sec:HHG in a homogeneous density medium}
In the case of a loose focusing geometry for HHG, the Rayleigh range is chosen to be long, typically in the order of tens of millimeters such that it becomes much longer than the length of the medium ($z_R\gg{L_{med}}$). Although loose focusing reduces the drive laser intensity, the advantage is that a large interaction volume can be provided for maximizing the total harmonic pulse energy. Usually a relatively low gas pressure is required to maintain phase matching~\cite{Midorikawa2008, Hergott2002, Rudawski2013, Tamaki2002}. In this geometry, the drive laser intensity stays approximately constant along the entire generating medium. Therefore, the harmonic dipole amplitude, $d_q(z,t)$, the interaction area, $S(z)$, and also the ionization fraction, $\eta(z,t)$, become space independent. Due to the long Rayleigh range, the geometry-induced wave-vector contribution can be simplified as $\Delta{k_{geo}}=-(q-1)/z_R$ and becomes independent of $z$. In addition, the harmonic dipole phase contribution, $\Delta{k_{dipole}}(z,t)$, becomes much smaller than the other terms~\cite{Kazamias2003-2} and can be neglected. As a result, the total wave-vector mismatch can be regarded as a space-independent parameter, $\Delta{k}(t)$, such that the phase term, $\phi_q(z,t)$, in Eq. \eqref{eq: instantaneous HHG field} can be simply represented as the product of $\Delta{k}(t)$ and $z$:
\begin{equation}
\label{eq: coherence length}
\phi_q(z,t)=\Delta{k}(t)\cdot{z}=\frac{\pi{z}}{L_{coh}(t)}.
\end{equation}
Here, the time-dependent coherence length, $L_{coh}(t)=\pi/{\Delta{k}(t)}$, is introduced for characterizing the wave-vector mismatch in comparison to other length parameters, specifically the length of the medium. The physical meaning of $L_{coh}$ is that the $q^{\rm{th}}$ harmonic radiation generated at the end of this length is shifted out of phase with respect to the harmonic radiation generated at the beginning of the length. This leads to back conversion in further propagation, often also termed as destructive inteference. By implementing a time-dependent coherence length, Eq. \eqref{eq: instantaneous HHG field} can be integrated analytically with the instantaneous HH intensity being~\cite{kazamias2011,Constant1999}
\begin{multline}
I_{q}(t)\propto\rho^2\left[1-\eta(t)\right]^2d_q(t)^2\frac{4L_{abs}^2}{1+4\pi^2\left[L_{abs}^2/L_{coh}(t)^2\right]}\\
\times\left[1+\mathrm{exp}\left(-\frac{L_{med}}{L_{abs}}\right)-2\mathrm{cos}\left(\pi\frac{L_{med}}{L_{coh}(t)}\right)\mathrm{exp}\left(-\frac{L_{med}}{2L_{abs}}\right)\right].
\label{eq: anaylitical solution of high harmonic field}
\end{multline}
In this expression, reabsorption of harmonic radiation is taken into account via an absorption length, $L_{abs}$. From Eq. \eqref{eq: anaylitical solution of high harmonic field} it can be seen that, ideally, the maximum harmonic output can be obtained when perfect phase matching is achieved, namely, $\Delta{k}(t)=0$ (equivalent to an infinitely long coherence length) during the entire drive laser pulse. In particular, when absorption is neglected (infinitely long absorption length) the harmonic intensity would grow proportionally to the square of the medium length. For the experimentally more relevant case that phase matching is not ideally fulfilled ($\Delta{k}\neq0$) and can be approximated by a temporally constant and finite coherence length, i.e., when the ionization fraction does not change much vs. time, and when additionally absorption cannot be neglected, a coarse criterion, $L_{coh}>L_{med}>3L_{abs}$, has been proposed~\cite{Constant1999, Midorikawa2008,Rudawski2013}. If this criterion is fulfilled, the harmonics are efficiently generated and the output is mainly limited by reabsorption. Quantitatively, for instance, when $L_{coh}>5L_{abs}$ and $L_{med}>3L_{abs}$, the harmonic intensity reaches more than half of the maximum intensity of the ideal case ($L_{coh}\rightarrow\infty$, $L_{abs}\rightarrow\infty$).
However, when generating high-order harmonics with an increased intensity and appreciable gas density, the named criterion becomes inapplicable because the coherence length becomes markedly. In that case, the coherence length varies strongly on an ultrafast time scale due to a dramatic change of the ionization fraction of the medium during the drive pulse. Therefore, phase matching is only achieved in a very short time interval during the drive pulse. As a result, for obtaining a maximum harmonic output pulse energy, it remains questionable to what fraction and at what instance during the drive laser pulse the criterion should be fulfilled. 
A further complication is that the harmonic dipole amplitude, $d_{q}(t)$, is also a time-dependent parameter that will influence the HH output. For instance, even if the above length criterion is fulfilled for a certain time interval, the output pulse energy may still be far below optimum if $d_{q}$ is small during that interval. 

In view of these effects and to translate them properly into a situation aiming for quasi-phase matching in Sec. \ref{sec:HHG in a periodic density modulated medium}, it is important to analyze the time dependence of the coherence length as well as of the harmonic dipole amplitude over the entire drive laser pulse. Here, for definiteness, we take two harmonic orders for HHG in argon as an example, namely, a moderate harmonic order, the $21^{\rm{st}}$ harmonic ($\approx38$ nm, HH21), and a relatively high harmonic order, the $39^{\rm{th}}$ harmonic ($\approx20$ nm, HH39).
\begin{figure}[ht!]
  \centering
  \includegraphics[width=0.8\textwidth]{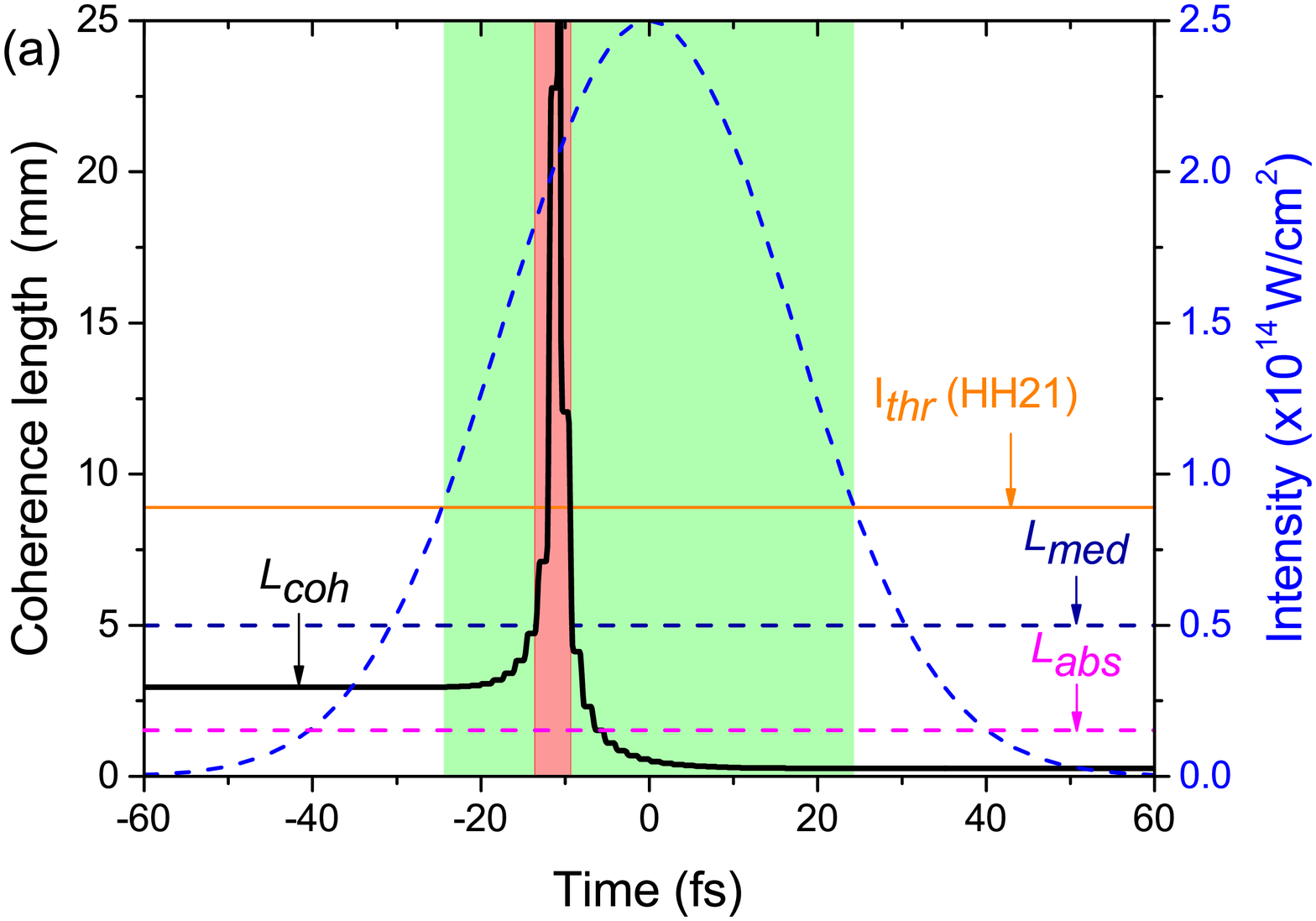}
  \includegraphics[width=0.8\textwidth]{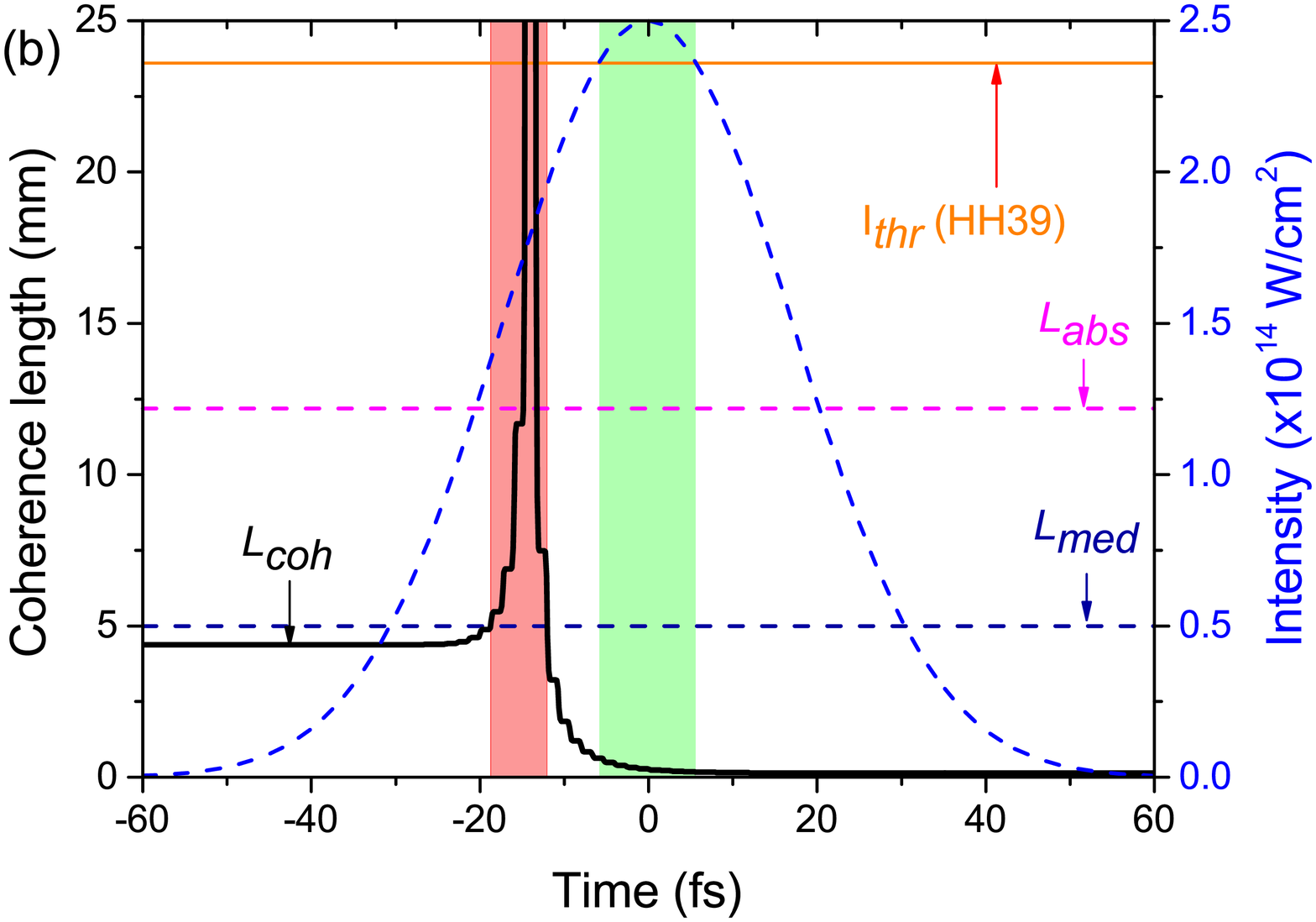}
  \caption{Calculated coherence lengths (black curves) for the  $21^{\rm{st}}$ harmonic (a) and the $39^{\rm{th}}$ harmonic (b) as a function of time. Assumed is harmonic generation in 30 mbar of argon in a 5 mm long medium ($L_{med}=5$ mm, dashed blue lines) with a Gaussian-shaped laser pulse of $2.5\times10^{14}\;\rm{W/cm^2}$ peak intensity (blue dashed curves). The absorption lengths (dashed pink lines) and the minimum required laser intensity (solid orange lines) are calculated for both harmonics and plotted for comparison. The green areas show within which time interval there is a sufficiently high drive laser intensity for HHG (${I>I_{thr}}$). The red areas indicate the time interval during which the phase-matching condition is approximately fulfilled, i.e., the harmonic output from the entire medium adds constructively ($L_{coh}>L_{med}$), if the respective harmonic is generated.}
 \label{fig:coherence length for 21st HH and 39th HH}
\end{figure}
Fig. \ref{fig:coherence length for 21st HH and 39th HH} shows the time-dependent coherence lengths (black curves) for the $21^{\rm{st}}$ harmonic (a) and the $39^{\rm{th}}$ harmonic (b) calculated for a HHG experiment in a typical loose focusing configuration, assuming a drive laser peak intensity of $I_0=2.5\times10^{14}\;\rm{W/cm^2}$, a Rayleigh range of $z_R$ =18 mm as achieved by a f/150 focusing lens, a medium length of $L_{med}=5$ mm and an argon gas pressure of 30 mbar. 
The laser pulse (also shown in Fig. \ref{fig:coherence length for 21st HH and 39th HH}) is again taken to have a Gaussian shape (blue dashed curve) with a pulse duration of $\tau=40\;\rm{fs}$ (FWHM). The lengths of the medium (dashed blue lines) and the calculated absorption lengths (dashed pink line) for both harmonics are also plotted for comparison. Furthermore, the green areas in Fig. \ref{fig:coherence length for 21st HH and 39th HH} (a) and (b) indicate the time interval of sufficiently high intensity in which, potentially, each respective harmonic can be generated ($I>I_{thr}$).

It can be clearly seen that for both harmonics, the coherence length shows an ultrafast time dependence. 
Specifically, starting from a certain mismatch, there is first a rapid increase of $L_{coh}$ in some part of the leading edge of the drive laser pulse. Initially, the dispersion of free electrons due to the negligible ionization fraction is relatively small, and the other terms in Eq. \eqref{eq: total phase mismatch} dominate. 
However, with the increase of the drive laser intensity, the ionization fraction rapidly increases further. Consequently, the dispersion in the plasma due to the accumulation of free electrons overwhelms the other terms in Eq. \eqref{eq: total phase mismatch}, resulting in a quick drop of the coherence length.
The short duration of the peak in the coherence length leaves only an extremely short interval (a few femtoseconds long) where the dispersion is close to zero (phase matching). 
For the $21^{\rm{st}}$ harmonic (Fig. \ref{fig:coherence length for 21st HH and 39th HH} (a)), the time interval of phase matched generation is indicated by the red area. This interval is found in the leading edge of the drive laser pulse (where the criterion $L_{coh}>L_{med}(5\;\textrm{mm})>3L_{abs}(0.7\;\rm{mm})$, is fulfilled). It can be seen that this interval overlaps with the time interval where the drive laser intensity is sufficiently high for generating a noticeable $21^{\rm{st}}$-order dipole moment, which is indicated as the green area. The overlap of the red with the green area indicates that the $21^{\rm{st}}$ harmonic can be efficiently generated at least within the short overlap time interval. 
For the $39^{\rm{th}}$ harmonic (Fig. \ref{fig:coherence length for 21st HH and 39th HH} (b)), the coherence length reaches a very large value and also becomes longer than the length of the medium in a short time interval at the leading edge of the pulse (red area). However, in contrast to the $21^{\rm{st}}$ harmonic, this interval is located outside the time interval in which the $39^{\rm{th}}$ harmonic can be generated (green area, where $I>I_{thr}(\textrm{HH39})$). Thus, the red (phase matching) and green (generation) intervals do not overlap, which indicates that the ionization fraction induced by the minimum required drive laser intensity for generating the $39^{\rm{th}}$ harmonic is too high for phase matching. The wave-vector mismatch, $\Delta{k}_{plasma}$, due to the large amount of free electrons dominates over the other wave-vector contributions. In this example, for argon, this occurs for harmonic orders higher than $q=35$.
Consequently, the $39^{\rm{th}}$ harmonic will not be efficiently generated at any moment throughout the entire drive laser pulse. 
\begin{figure}[ht!]
  \centering
  \includegraphics[width=0.7\textwidth]{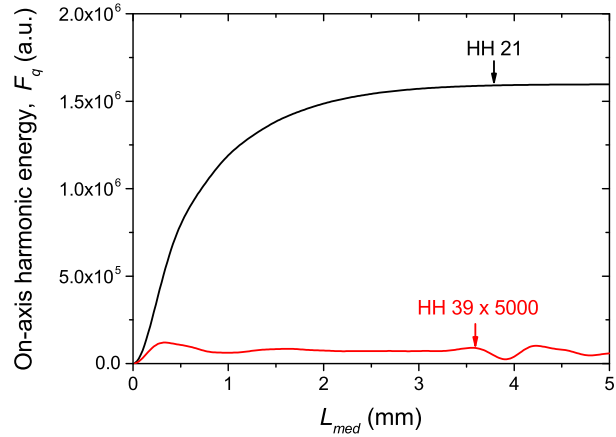}
  \caption{Calculated HH pulse energy for the 21$^{\rm{st}}$ (black) and the 39$^{\rm{th}}$ harmonic (red) as a function of the medium length. For $L_{med}>1\;\rm{mm}$, the 39$^{\rm{th}}$ harmonic output is about four to five orders of magnitude weaker.}
 \label{fig:HH intensity for 21st HH and 39th HH}
\end{figure}
Also, to quantitatively illustrate the difference between the two considered harmonics, we have calculated the on-axis harmonic output pulse energy, $F_q$, as a function of interaction length using Eqs. \eqref{eq: HHG intensity} and \eqref{eq: anaylitical solution of high harmonic field}. This is plotted in Fig. \ref{fig:HH intensity for 21st HH and 39th HH}. Note that the cross-sectional area, $S$, is treated as a constant and can be calculated from the waist of the drive laser beam at the focus.
It can be seen that the output energy of the $21^{\rm{st}}$ harmonic (black curve) grows strongly within the first few mm of the medium and saturates after roughly 3 to 4 mm. This observation shows that the harmonic is efficiently generated and only limited by reabsorption. On the other hand, the output pulse energy of the $39^{\rm{th}}$ harmonic (red curve, multiplied by 5000 in order to be visible in the graph) does not show any obvious growth with the length of the medium and is (beyond $L_{med}=1\;\rm{mm}$) approximately four to five orders of magnitude weaker. A few oscillations can be seen along the length of the medium, which is due to varying levels of destructive interference. 

From a series of additional calculations, we conclude that for moderate harmonic orders up to $q=35$ in this example, the length criterion for efficient output ($L_{coh}>L_{med}>3L_{abs}$) can be fulfilled over at least some certain short time interval falling within the interval that the drive laser intensity is above the generation threshold. This allows for harmonic generation that is limited only by absorption. However, for harmonics of $q>35$ (such as in the example above, $q=39$), the criterion can never be fulfilled at any instance during the drive laser pulse, which leads to a generation efficiency that is decreased by many orders of magnitude. 

A solution to overcome such low efficiencies for higher orders is to implement QPM techniques into the HHG configuration. In Section \ref{sec:HHG in a periodic density modulated medium}, we present a specific maximally simplified QPM scheme based on providing a periodic density modulation.

\section{HHG in a periodic density modulated medium}
\label{sec:HHG in a periodic density modulated medium}
The basic idea of quasi-phase matching (QPM) is to introduce some periodic modulation of the local strength or phase of the generation process throughout the medium. To counteract the destructive interference between locally generated harmonic fields, i.e., generated in subsequent coherence lengths, a straightforward approach would be to suppress the harmonic generation within every other coherence length. By repeating the suppression periodically in the spatial domain, such modulation is equivalent to providing an additional wave-vector contribution in the Fourier domain, $K_{QPM}$, that compensates the remaining wave-vector mismatch ($\Delta{k},(z,t)$ in Eq. \eqref{eq: total phase mismatch}). As was shown above for harmonics in argon with an order beyond $q=35$, the mismatch that needs to be compensated originates predominately from the plasma dispersion. The total wave-vector mismatch $\Delta{k'},(z,t)$, for HHG including QPM can be expressed as $\Delta{k'},(z,t)=\Delta{k},(z,t)-K_{QPM}$. 
\begin{figure}[ht!]
  \centering
  \includegraphics[width=0.8\textwidth]{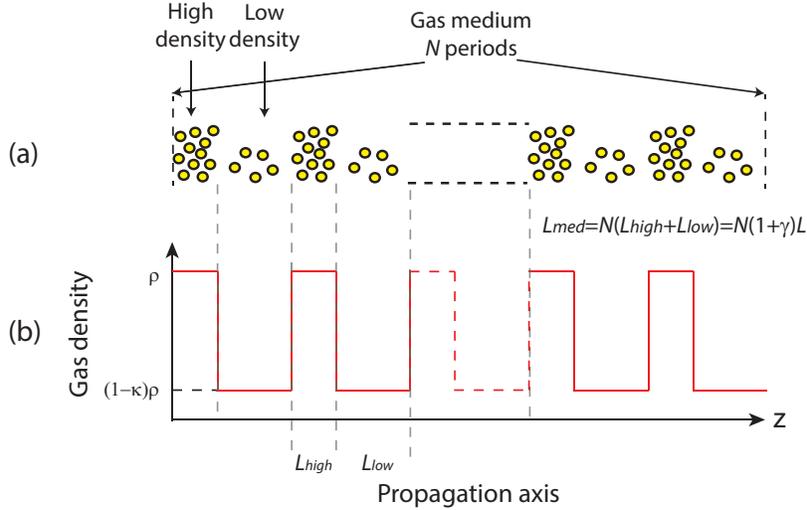}
  \caption{(a) Schematic of a density modulated medium consisting of alternating regions with high and low gas density. (b) The corresponding gas density profile can be described via a maximum gas density, $\rho$, and a modulation depth, $\kappa$. The lengths of the high-density and low-density regions are $L_{high}$ and $L_{low}$, respectively.}
 \label{fig: density modulated medium}
\end{figure}
The latter expression looks simple. However, two points have to be noted. While the mismatch $\Delta{k}(z,t)$ changes on an ultrafast time scale, the additional wave-vector, $K_{QPM}$, essentially has a temporally constant value in the case of using a periodic density modulated medium because it is impossible to change any gas density distribution on an ultrashort time scale as well. Consequently, quasi-phase matching will only be fulfilled transiently again, for a certain short time interval within the drive laser pulse, as discussed above. Moreover, because the density of the medium is spatially modulated, the mismatch attains an additional, periodic space dependence.

We will consider here a periodic modulation of the gas density along the propagation axis, as is depicted in Fig. \ref{fig: density modulated medium}, where we assume a rectangular variation for simplicity. Similar density distribution might be achieved, e.g., by placing a periodic grid in a supersonic gas jet. To express the density modulation quantitatively, the length of the medium can be written as: $L_{med}=NP_{mod}=N(L_{high}+L_{low})$, where $N$ is the number of modulation periods, $P_{mod}$ is the modulation period, and $L_{high}$ and $L_{low}$ are the lengths of the high-density and low-density regions, respectively.
The corresponding modulated gas density profile along the propagation axis of the drive laser, $z$, is shown in Fig. \ref{fig: density modulated medium} (b) using the simplifying assumption of homogeneous densities within $L_{high}$ and $L_{low}$. We specify the modulation depth of the gas density with a factor $\kappa$. When $\kappa$=1, the gas density is fully modulated, while if $\kappa$=0, there is no modulation imposed on the gas. To obtain the instantaneous on-axis harmonic field, $E_q(t)$, in a density modulated medium with a loose focusing geometry, one has to take into account that several additional parameters become space dependent in the integral in Eq. \eqref{eq: instantaneous HHG field}, specifically, the gas density, $\rho(z)$, the phase term, $\phi_q(z,t)$ and the absorption length, $L_{abs}(z)$. To evaluate Eq. \eqref{eq: instantaneous HHG field}, we rewrite the integral as a sum over the field contributions from the high-density and low-density regions as
\begin{equation}
\label{eq: sum of the instantaneous harmonic field}
E_{q}(t)=\sum^N_{n=1}\left[E_{q,n}^{high}(t)+E_{q,n}^{low}(t)\right],
\end{equation}
where $E_{q,n}^{high}(t)$ and $E_{q,n}^{low}(t)$ are the $q^{\rm{th}}$ harmonic field contributions generated in the high-density and low-density regions of the $n^{\rm{th}}$ modulation period, respectively. The phase term, $\phi_{q,n}(z,t)$, can be calculated from the wave-vector mismatch in the regions of different densities as follows
\begin{equation}
\label{eq: Phi}
\phi_{q,n}(z,t)=\left\{
\begin{aligned}
(n-1)\left[\Delta{k}_{high}(t)L_{high}+\Delta{k}_{low}(t)L_{low}\right]\\
+\Delta{k}_{high}(t)\cdot{z}\;\;\;\;\;\;\;\;\rm{in\;any\;high{-}density\;region}, \\
(n-1)\left[\Delta{k}_{high}(t)L_{high}+\Delta{k}_{low}(t)L_{low}\right]\\
+\Delta{k}_{high}(t)L_{high}+\Delta{k}_{low}(t)\cdot{z}\;\;\;\;\;\;\;\; \rm{in\;any\;low{-}density\;region}.
\end{aligned}
\right.
\end{equation}
Here, $\Delta{k}_{high}(t)$ and $\Delta{k}_{low}(t)$ is the wave-vector mismatch in the high-density and low-density regions, respectively. Similarly, the coherence length in the high-density and low-density regions can be written as: $L_{coh}^{high}(t)=\pi/\Delta{k}_{high}(t)$ and $L_{coh}^{low}(t)=\pi/\Delta{k}_{low}(t)$. 

Next, in order to investigate the influence of QPM on efficient harmonic generation rather than on limitations due to reabsorption, we restrict ourselves to the situation that the length of the medium is shorter than the absorption length. As was described in Sec. \ref{sec:HHG in a homogeneous density medium}, this can be achieved with a sufficiently low gas pressure in a loose focusing geometry. Indeed, the absorption lengths for the harmonics of higher-valued order ($q>35$) are typically much longer than the length of the medium (5 mm) over a broad tuning range of the gas pressure (density). For example, the absorption length for the $39^{\rm{th}}$ harmonic at a pressure of 30 mbar is about 12 mm long even at non-modulated gas density.

Using the equation for the sum of a geometric series, Eq. \eqref{eq: sum of the instantaneous harmonic field}, the expression for the instantaneous on-axis harmonic field can be simplified, and reads
\begin{equation}
E_q(t)=\xi_{q}^{high}(t)+\xi_{q}^{low}(t),
\end{equation}
where $\xi_{q}^{high}(t)$ is the sum of generated harmonic field contributions from only the high-density regions:
\begin{equation}
\xi_{q}^{high}(t)\propto\frac{\rho{d_q(t)}}{i\pi/L_{coh}^{high}(t)}\left[\mathrm{exp}\left(i\frac{\pi{L_{high}}}{L_{coh}^{high}(t)}\right)-1\right]\frac{1-\mathrm{exp}\left[iN\pi\left(\frac{L_{high}}{L_{coh}^{high}(t)}+\frac{L_{low}}{L_{coh}^{low}(t)}\right)\right]}{1-\mathrm{exp}\left[i\pi\left(\frac{L_{high}}{L_{coh}^{high}(t)}+\frac{L_{low}}{L_{coh}^{low}(t)}\right)\right]},
\label{eq: E high}
\end{equation}
and where $\xi_{q}^{low}(t)$ is the generated harmonic field contribution from only the low-density regions:
\begin{multline}
\xi_{q}^{low}(t)\propto\frac{(1-\kappa)\rho{d_q(t)}}{i\pi/L_{coh}^{low}(t)}\left[\mathrm{exp}\left(i\frac{\pi{L_{low}}}{L_{coh}^{low}(t)}\right)-1\right]\mathrm{exp}\left[i\frac{\pi{L_{high}}}{L_{coh}^{high}(t)}\right]\\
\times\frac{1-\mathrm{exp}\left[iN\pi\left(\frac{L_{high}}{L_{coh}^{high}(t)}+\frac{L_{low}}{L_{coh}^{low}(t)}\right)\right]}{1-\mathrm{exp}\left[i\pi\left(\frac{L_{high}}{L_{coh}^{high}(t)}+\frac{L_{low}}{L_{coh}^{low}(t)}\right)\right]}.
\label{eq: E low}
\end{multline}
With these expressions, the instantaneous on-axis intensity, $I_{q}(t)$, for the $q^{\rm{th}}$ harmonic from a density modulated medium becomes
\begin{equation}
I_{q}(t)=\frac{1}{2}c\varepsilon_0|\xi_{q}^{high}(t)+\xi_{q}^{low}(t)|^2.
\label{eq: QPM instantaneous HH intenisity}
\end{equation}
The corresponding on-axis HH pulse energy of the $q^{\rm{th}}$ harmonic, $F_q$, can be obtained again by integration over the entire pulse (Eq. \eqref{eq: HHG intensity}). 

From Eqs. \eqref{eq: E high} and \eqref{eq: E low}, it can be seen again that the coherence length (in both the high-density and low-density regions, $L_{coh}^{high}$ and $L_{coh}^{low}$) varies strongly on an ultrafast time scale via the ionization fraction. Therefore, similar to standard phase matching discussed above for lower-order harmonics, QPM can only be transiently achieved in a short time interval within the drive laser pulse. However, by choosing the value of the modulation period, this time interval can be shifted to a proper instance, such that the HH pulse energy is maximized.

\section{Results and Discussion of HHG with QPM}
\label{sec: results and discussion}
In this section, we describe quasi-phase matched generation of the $39^{\rm{th}}$ harmonic ($\approx20$ mm). We consider this specific harmonic again, because it represents the typical case where standard phase matching cannot be achieved, as was described in Sec. \ref{sec:HHG in a homogeneous density medium}. We assume the same loose focusing parameters given in Sec. \ref{sec:HHG in a homogeneous density medium} (a drive laser peak intensity $I_0=2.5\times10^{14}$ $\rm{W/cm^2}$, a Rayleigh range $z_R$ =18 mm achieved by a f/150 focusing lens, $L_{med}=5$ mm) and the same argon gas pressure (30 mbar). Based on the QPM model described in Sec. \ref{sec:HHG in a periodic density modulated medium}, we calculate the instantaneous on-axis HH intensity, $I_{\rm{39}}(t)$, and the harmonic pulse energy, $F_{\rm{39}}$, by applying different modulation periods, $P_{mod}$, and modulation depths, $\kappa$. 
We note that, to re-phase the harmonic wave with the fundamental wave after each coherence length in the high-density regions, the low-density regions have to provide a phase shift of $\pi$ between the generated harmonic wave and the fundamental wave. This means that also the low-density regions have to be as long as a coherence length, which is dependent of the modulation depth, $\kappa$. As a result, the desired modulation period should be equal to the sum of the coherence lengths calculated in both high-density and low-density regions, $P_{mod}=L_{coh}^{high}+L_{coh}^{low}$. 

In order to choose an appropriate modulation period, $P_{mod}$, the time-dependent coherence length for both high-density and low-density regions needs to be calculated.
\begin{figure}[ht!]
\centering
  \includegraphics[width=0.8\textwidth]{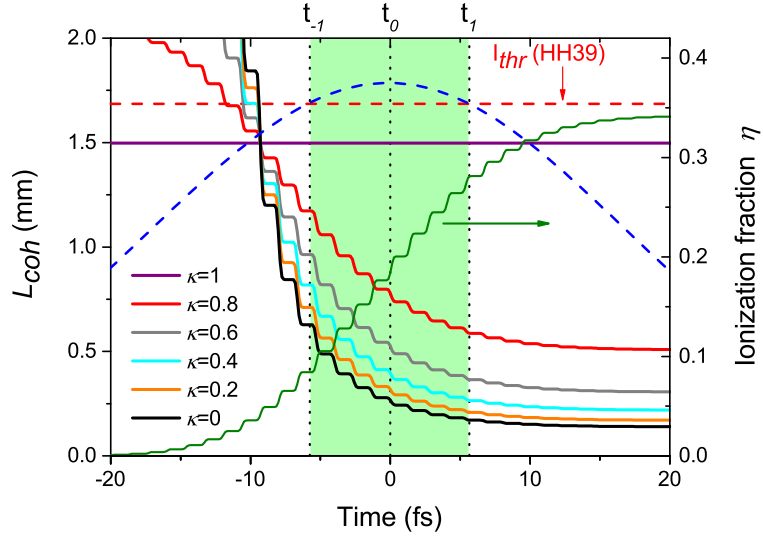}
  \caption{Calculated coherence lengths in the low-density region for the $39^{\rm{th}}$ harmonic ($\approx20$ mm) as a function of time for various modulation depths, $\kappa$ (1 (purple), 0.8 (red), 0.6 (grey), 0.4 (cyan), 0.2 (orange) and 0 (black)). The ionization fraction (dark green curve) and the temporal profile of the drive laser intensity (blue dashed curve) are plotted for reference. The red dashed line is the threshold intensity for the 39$^{\rm{th}}$ harmonic generation and the green area indicates the time interval (from ${t_{-1}}$ to ${t_{1}}$) during which the generation of the $39^{\rm{th}}$ harmonic is possible. The generating medium is 5 mm long and the gas pressure is  set at 30 mbar in the high-density region. The drive laser pulse has a peak intensity of $2.5\times10^{14}\;\rm{W/cm^2}$.}
  \label{fig:different coherence lengths for 39HH}
\end{figure}
Fig. \ref{fig:different coherence lengths for 39HH} shows the calculated coherence lengths  for the $39^{\rm{th}}$ harmonic in the low-density regions for a set of values ($\kappa>0$) and high-density region ($\kappa=0$) as a function of time. The modulation depth, $\kappa$, is varied from 0 to 1 in steps of 0.2. The drive laser pulse is shown as the blue dashed curve.
The green area indicates again the time interval where the drive laser intensity is above the intensity threshold (red dashed line) for generating the $39^{\rm{th}}$ harmonic. 

In Fig. \ref{fig:different coherence lengths for 39HH}, for easier discussion, three characteristic times are introduced and shown along the top axis, where $t_{-1}$ represents the on-set of the generation of the $39^{\rm{th}}$ harmonic, $t_{0}$ is the moment at which the drive laser pulse reaches its peak intensity, and $t_{1}$ indicates the time when the drive laser intensity drops below the intensity threshold for generating the $39^{\rm{th}}$ harmonic. For full modulation ($\kappa=1$, purple line), the coherence length for the low-density region becomes independent of time due to the absence of any gas that can be ionized. In this limiting case, the coherence length is only given by the wave-vector mismatch, $\Delta{k_{geo}}$, due to the Gouy phase shift, which is set by the focusing geometry. The coherence lengths for the modulation depths with $\kappa<1$ are strongly time-dependent and drop rapidly during the pulse due to the increasing ionization fraction (dark green curve), specifically within the time interval where the $39^{\rm{th}}$ harmonic is generated (green area).
The information in the figure can be used as follows: if the harmonic output is to be maximized for a particular case, one can read the required QPM period from the figure. For instance, if the output is to be maximized in a short time interval at the peak of the drive laser pulse, the crossing of the coherence lengths with the dashed vertical line at that time ($t_0$) gives the length of both the high-density and low-density regions that yields QPM at that time. For instance, for $\kappa=0.8$, one obtains $L^{low}_{coh}=0.79$ mm (red curve) and $L^{high}_{coh}=0.27$ mm (black curve), resulting in a modulation period of $P_{mod}=L_{coh}^{high}+L_{coh}^{low}=1.06$ mm. 

\begin{figure}[ht!]
\centering
\includegraphics[width=0.62\textwidth]{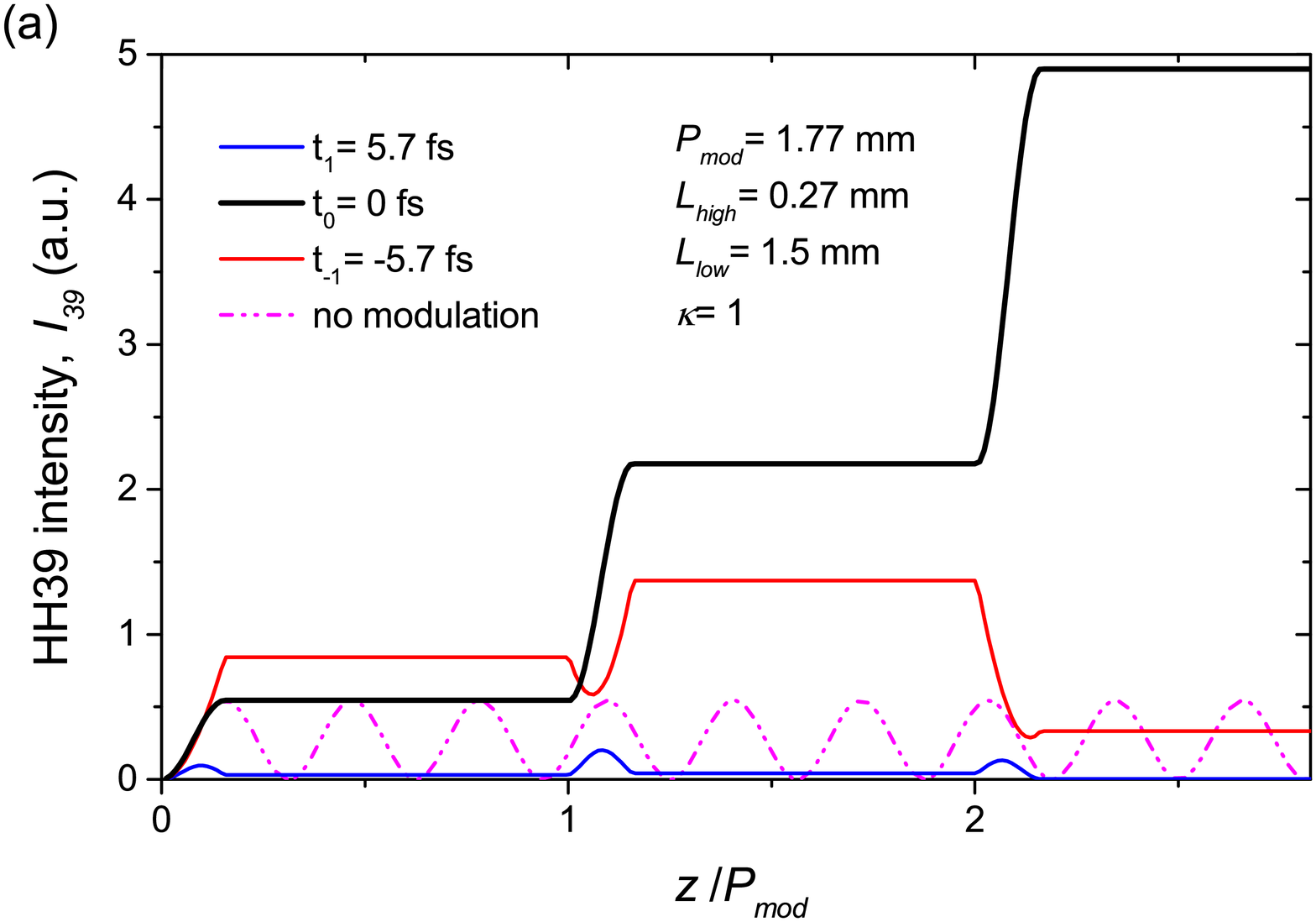}
\includegraphics[width=0.62\textwidth]{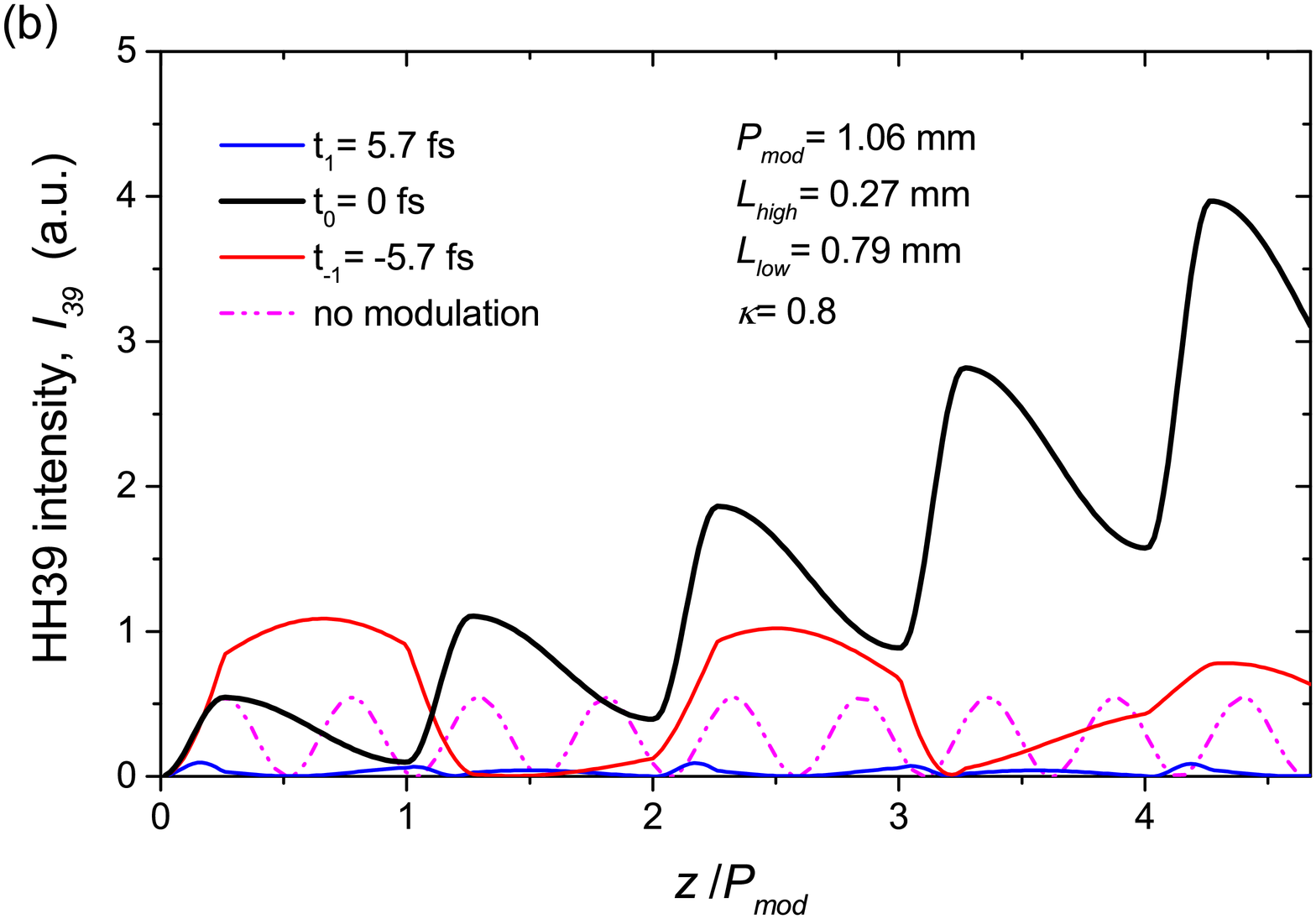}
\includegraphics[width=0.62\textwidth]{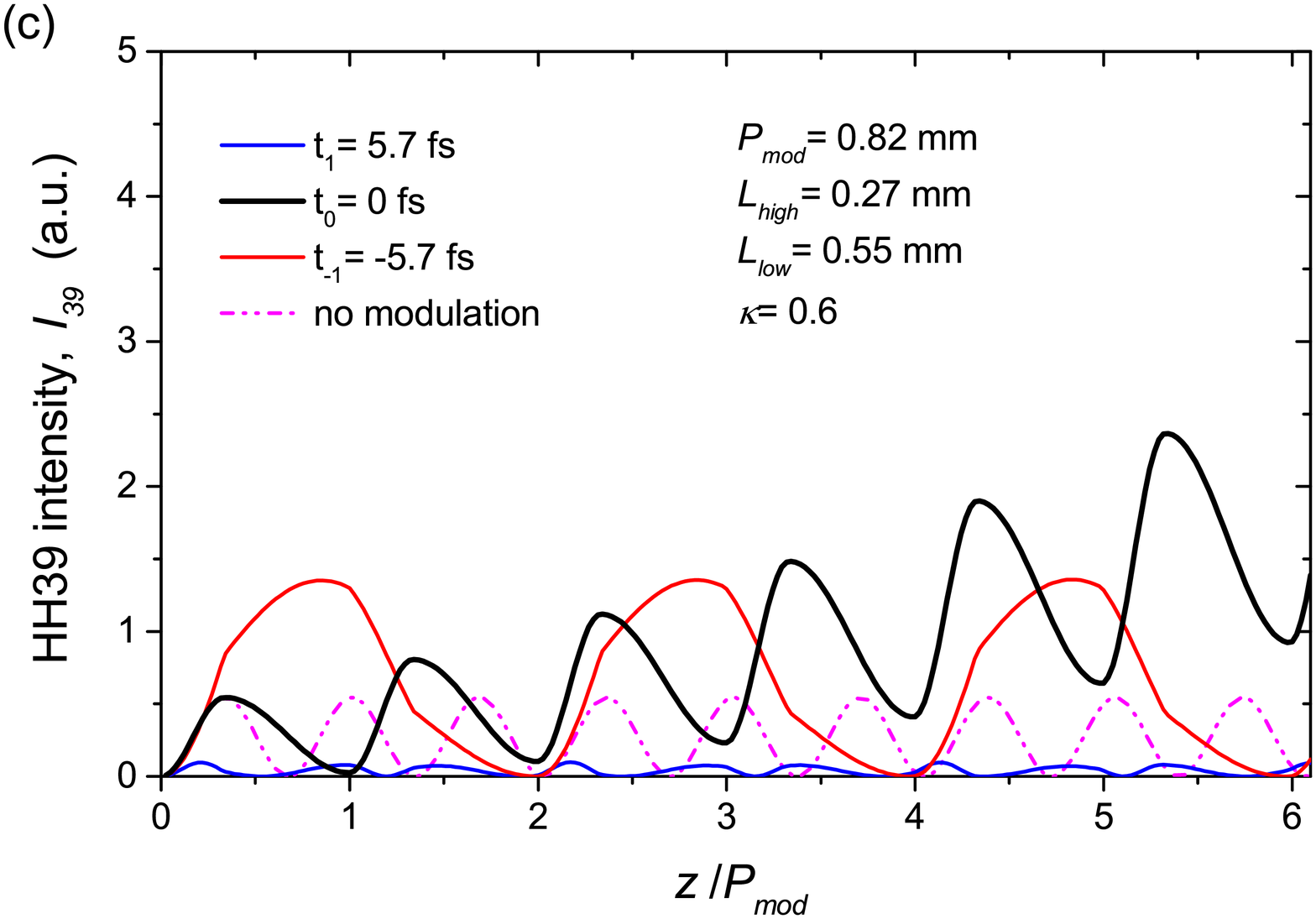}
\caption{Growth of the on-axis HH intensity, $I_{\rm{39}}$, throughout the density modulated medium for (a) $\kappa$=1, (b) $\kappa$=0.8 and (c) $\kappa$=0.6 when the modulation period, $P_{mod}$, is chosen to induce QPM at the peak of the drive laser pulse ($t_0$=0 fs). The black curves show the growth of $I_{\rm{39}}$ generated at the drive laser peak, while the other curves show the growth of  $I_{\rm{39}}$ at an earlier and later time in the pulse ($t_1$=-5.7 fs (red) and $t_3$=5.7 fs (blue)). The propagation coordinate is expressed relative to the modulation period. The pink dashed line represents $I_{39}$ generated at $t_0$ in a homogeneous medium, without density modulation ($\kappa$=0).}
  \label{fig:HH39 output at different time}
\end{figure}

Now that we have illustrated how to choose the modulation period that maximizes $I_q$ in Eq. \eqref{eq: QPM instantaneous HH intenisity} at the peak of the drive laser pulse, $t_0$, we present in Fig. \ref{fig:HH39 output at different time} a calculation of how the HH intensity ($I_{\rm{39}}$) generated at the drive laser peak grows vs. the propagation coordinate through the medium, $z$. For comparison, we also plot the spatial growth of $I_{\rm{39}}$ generated at other characteristic instances of the drive laser pulse in the leading edge ($t_{-1}=$-5.7 fs) and the trailing edge ($t_{1}=$5.7 fs).
With the modulation period, $P_{mod}$, set to provide QPM at $t_0$, we also display the influence of the modulation contrast by choosing three different modulation depths.
For reference, we also plot the growth of $I_{\rm{39}}$ generated at the peak of the drive laser pulse in a homogeneous medium, i.e., without density modulation. 

From Fig. \ref{fig:HH39 output at different time} it can be clearly seen that, in the absence of density modulation, the harmonic intensity generated at $t_0=$0 fs does not grow vs. the interaction length but only oscillates periodically between 0 and some smaller value.
This oscillation, indicating periodic generation and back conversion, is caused by a rather short coherence length (in this case about 270 $\mu{\rm{m}}$). However, when the medium is periodically modulated, the harmonic intensity of the $39^{\rm{th}}$ harmonic generated at $t_0$=0 fs (black curves) shows a clear growth of the intensity with the medium length with all shown examples of the modulation depth. Specifically, the intensity generated at $t_0$ increases in synchronism with the increasing number of modulation periods. Such increase of the intensity implies that QPM is successfully achieved due to a correctly chosen modulation period. However, the harmonic intensity generated at the other times of the drive pulse (red and blue curves) oscillates between 0 and very low values along the medium length because for these generation times the modulation period is incorrectly chosen. Among the black curves in Fig. \ref{fig:HH39 output at different time} ($t=t_0$), the highest harmonic intensity is found in Fig. \ref{fig:HH39 output at different time} (a), where a full modulation is applied ($\kappa$=1). This can be explained as follows: due to the absence of gas in the low-density regions, a full suppression of the out-of-phase generation is achieved. For other modulation depths, the $39^{\rm{th}}$ harmonic generated from the remaining gas in the low-density region will still partially interfere destructively with the harmonics generated in the high-density region. Moreover, it can be noted that the maximum $I_{\rm{39}}$, generated earlier in the pulse (at $t_{-1}$, red curves) is always higher than that generated with the same intensity later in the pulse (at $t_{1}$, blue curves), although the harmonic dipole amplitude, $d_{\rm{39}}$, is the same at both times. This is due to a larger number of neutral atoms contributing to the HHG in the leading edge of the pulse than that in the trailing edge, because the ionization fraction reduces the neutral atomic density. 

\begin{figure}[ht!]
  \centering
  \includegraphics[width=0.8\textwidth]{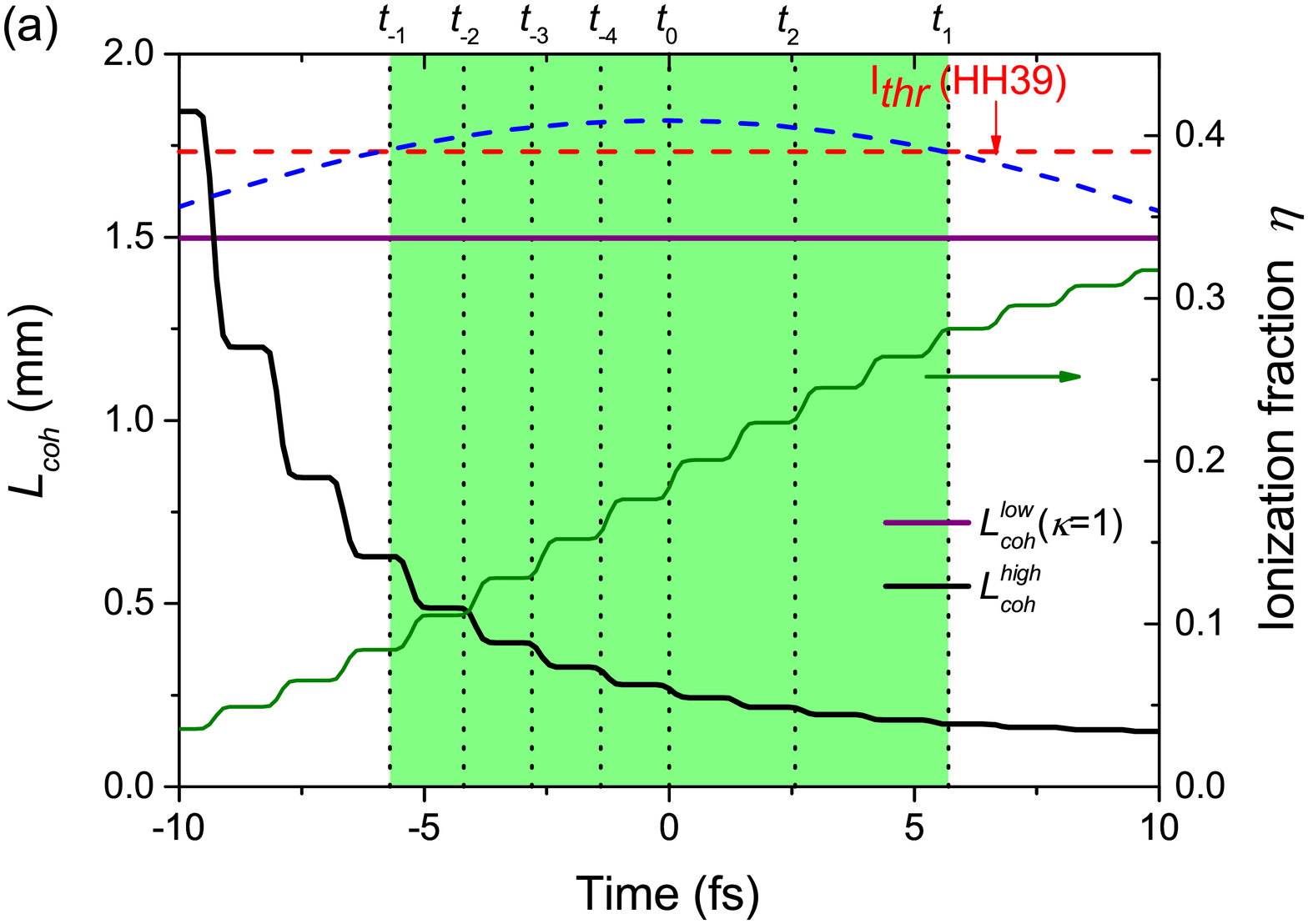}
  \includegraphics[width=0.8\textwidth]{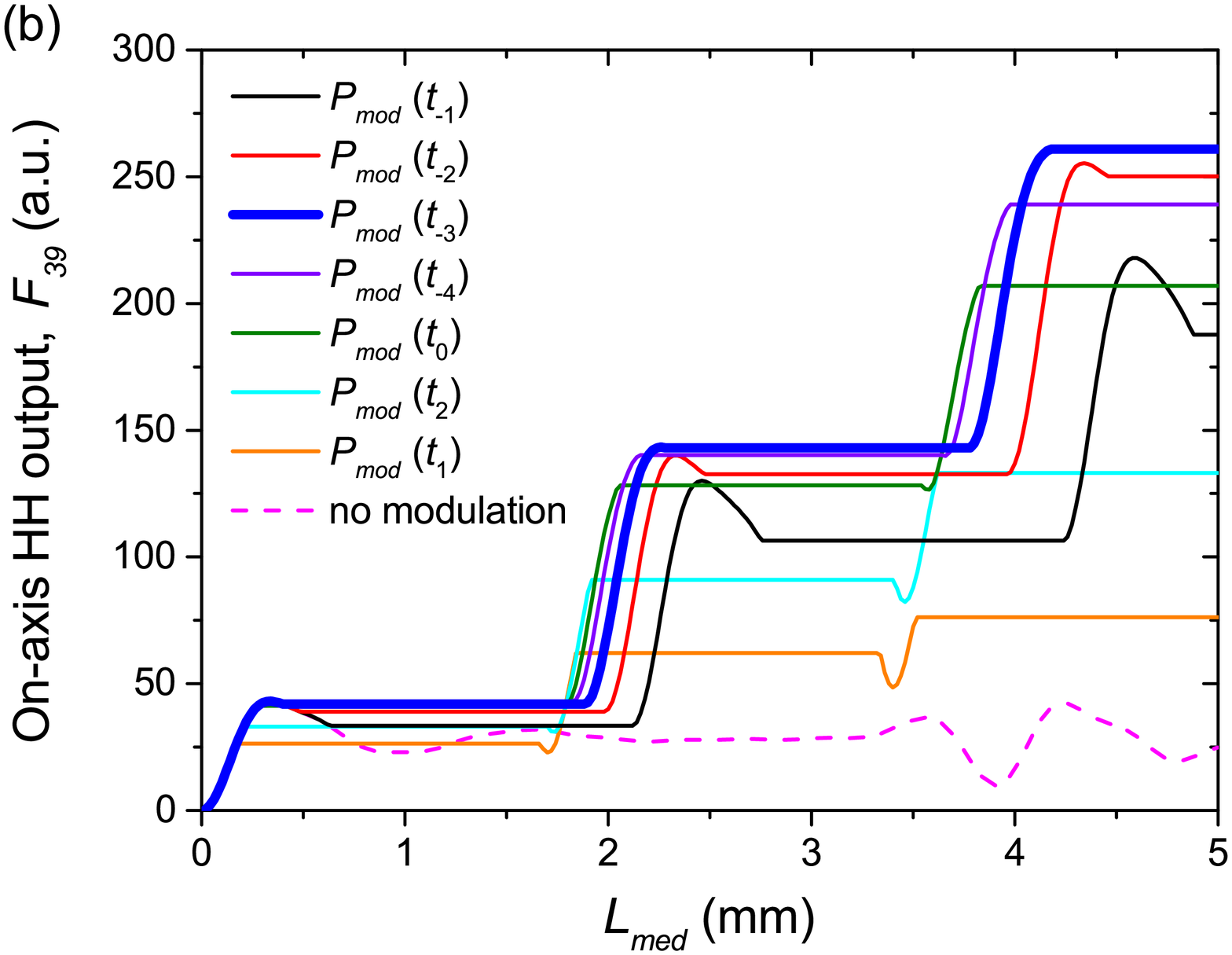}
  \caption{(a) calculated coherence lengths enlarged from Fig. \ref{fig:different coherence lengths for 39HH}. The purple horizontal line is the coherence length in the low-density regions, while the black curve is the coherence length in the high-density regions. The modulation depth is set to $\kappa$=1. For the calculation of the HH output pulse energy, seven different modulation periods are chosen corresponding to providing QPM at seven selected instances in the drive laser pulse (from $t_{-1}$ to $t_{1}$, dotted lines). (b) On-axis HH pulse energy for the $39^{\rm{th}}$ harmonic versus the length of the medium obtained with the seven different modulation periods. The highest output is achieved with $P_{mod}$ chosen  to provide transient QPM in the rising edge of the pulse ($t_{-3}$, thick blue line). The pink dashed line of the lowest output is obtained without density modulation.}
  \label{fig:total HH39 output}
\end{figure}

The results above show that choosing a particular modulation period only maximizes the HH output for a certain temporal fraction of the drive laser pulse, but, that also at other time intervals there is output, particularly in the leading edge.
In an experiment, however, one is usually observing the HH pulse energy, while the instantaneous HH intensity is rather difficult to monitor~\cite{Paul2011}. To optimize the pulse energy, $F_{q}$, we integrate the instantaneous harmonic intensity, $I_q(t)$, over the entire drive laser pulse and vary the modulation period. Since, from the previous calculations, the strongest growth in the instantaneous harmonic intensity is found in a fully density modulated medium ($\kappa$=1), we restrict ourselves to this example. In previous experiments, generally a modulation period has been selected that is equal to the sum of the coherence lengths calculated for high-density and low-density regions at the time that the drive laser pulse reaches the peak intensity. The motivation for such choice was that at that time the highest harmonic dipole amplitude, $d_{\rm{39}}$, can be achieved~\cite{Hage2014, Gibson2003, Liu2012OL}. However, this choice appears questionable because then the growth of the harmonic intensity is placed in the time of the highest ionization rate (quickest change of plasma dispersion) vs. time which yields the shortest possible time interval of quasi-phase matching. 
On the other hand, if QPM is set to occur over a longer time interval (when the ionization rate is not maximum), one has to take into account a smaller instantaneous intensity.
To determine the optimal modulation period for the highest on-axis harmonic pulse energy, we again apply the same loose focusing configuration for the experiment described in Sec. \ref{sec:HHG in a homogeneous density medium}. We calculate the harmonic pulse energy for the $39^{\rm{th}}$ harmonic by selecting different modulation periods that provide QPM at various different time intervals of sufficiently high intensity (green area in Fig. \ref{fig:different coherence lengths for 39HH}). For more clarity, we have replotted the green area on a finer time scale in Fig. \ref{fig:total HH39 output} (a). 
It can be seen that between $t_{-1}$ and $t_1$, the coherence length in the low-density regions ($L_{coh}^{low}$) remains independent of time due to the absence of gas that can be ionized, while the coherence length in the high-density regions ($L_{coh}^{high}$) decreases due to the increasing ionization fraction.
Here, we choose various different modulation periods that provide QPM at seven different instances in the drive pulse: at five instances in the leading edge, denoted as $P_{mod}(t_{-1})$, $P_{mod}(t_{-2})$, $P_{mod}(t_{-3})$, $P_{mod}(t_{-4})$ and $P_{mod}(t_{0})$, respectively, and at another two instances in the trailing edge, denoted as $P_{mod}(t_{2})$ and $P_{mod}(t_{1})$. The calculated lengths for the high-density and low-density regions as well as the corresponding modulation periods are listed in Tab. \ref{tb: modulation period}.
\begin{table}[ht!]
\centering
\caption{List of coherence lengths in the high-density and low-density regions and the modulation periods ($P_{mod}$) chosen to provide QPM at seven different instances within the drive laser pulse when the 39$^{\rm{th}}$ harmonic is generated.}
\begin{tabular}{|c|c|c|c|}
\hline
 & $L_{high}$ (mm) & $L_{low}$ (mm) & $P_{mod}$ (mm) \\ \hline
$t_{-1}$ =-5.7 fs   & 0.63 & 1.50 & 2.13 \\ \hline
$t_{-2}$ =-4.2 fs   & 1.98 & 1.50 & 1.98 \\ \hline
$t_{-3}$ =-2.8 fs   & 0.39 & 1.50 & 1.89 \\ \hline
$t_{-4}$ =-1.4 fs   & 0.32 & 1.50 & 1.82 \\ \hline
$t_{0}$ = 0   fs   & 0.27 & 1.50 & 1.77 \\ \hline
$t_{2}$ = 2.8 fs   & 0.20 & 1.50 & 1.70 \\ \hline
$t_{1}$ = 5.7 fs   & 0.17 & 1.50 & 1.67 \\
\hline
\end{tabular}
\label{tb: modulation period}
\end{table}
In Fig. \ref{fig:total HH39 output} (b), we show the calculated on-axis harmonic pulse energy, $F_{\rm{39}}$, versus the propagation coordinate through the medium for the listed modulation periods. For comparison with the case where QPM is absent, we calculated the HH pulse energy from a homogeneous medium as well.
It can be seen that the output pulse energy is very sensitive to the chosen modulation period.

For a homogeneous medium (without density modulation, pink dashed curve), the output oscillates along the medium length and only reaches a small value. In the density modulated medium, the output grows stepwise along the propagation direction with each modulation period, which indicates successful QPM in spite of its transient nature.
The most important finding from Fig. \ref{fig:total HH39 output} (b) is that, due to this transient nature, the highest on-axis harmonic pulse energy is obtained by applying a modulation period that provides transient QPM in the \emph{leading edge} rather than at the peak intensity of the drive laser pulse, here with $P_{mod}(t_{-3})$. In this example with  the $39^{\rm{th}}$ harmonic in argon, the optimum QPM period is about 50\% bigger than that for the peak of the laser pulse and provides 25\% more output pulse energy. 

In summary, the optimum modulation period for the highest on-axis HH pulse energy, $F_{\rm{39}}$, can only be approximately be determined from the remaining wave-vector mismatch at the peak of the drive laser pulse. Although at this time the highest instantaneous output intensity is reached and even higher output is expected with larger QPM period. We assign this to the fact that before the peak the time interval of successful QPM longer and overcompensates a some what lower intensity in the rising edge.

\section{Conclusion}
We have developed a one-dimensional, time-dependent model for quasi-phase matching (QPM) of high-order harmonic generation. The model is used to calculate the optimum quasi-phase-matching period, taking into account the highly dynamic change of index during harmonic generation. The model is presented using an example where standard phase matching cannot be achieved, here the $39^{\rm{th}}$ harmonic to be generated in argon. Using a spatially periodic density modulation (periodic structure), the quasi-phase matching occurs transiently at some instance during the drive laser pulse. This is due to strong ionization driven by the need for high intensities and therefore leading to an ultrafast dependence of the coherence length. The choice of the modulation period determines the specific moment and duration of quasi-phase matching. We show that simply choosing the modulation period to provide phase matching at the peak intensity of the pulse, when the highest amplitude of the nonlinearly induced high-order harmonic dipole amplitude is reached, does not provide the highest harmonic output pulse energy. To find the optimum modulation period, it is essential to analyze the time dependence of the wave-vector mismatch and the harmonic dipole amplitude along the entire pulse duration. According to our model, the optimal modulation is obtained with longer periods than usually assumed, such that transient quasi-phase matching is provided in the leading edge of the pulse. 

Although we have presented results from the model using a specific choice of parameters for an idealized situation, we note that the model can also be applied to more complicated modulation profiles such as would be present e.g., in a density modulated cluster jet. The model may also be applied to more complicated gas distributions such as multi dual-gas jet arrays~\cite{Willner2011PRL}, by recalculating the wave-vector mismatch and the induced dipole amplitudes for the different gas species used. Further improvements of this model would include an extension to three dimensions (including the transverse direction) together with an improved quantum mechanical calculation of time-dependent single-atom response compared to the power-law approximation used in this paper for determining a more accurate QPM period.

\section*{Acknowledgement}
This research was supported by the Dutch Technology Foundation STW, which is part of the Netherlands Organization for Scientific Research (NWO), and partly funded by the Ministry of Economic Affairs (Project No. 10759).
\end{document}